\documentstyle[12pt]{article}
\input epsf
\newcommand{\beq}{\begin{equation}}

\newcommand{\beqarr}{\begin{eqnarray}}
\newcommand{\eeqarr}{\end{eqnarray}}

\newcommand{\be}{\begin{equation}}
\newcommand{\ee}{\end{equation}}

\newcommand{\ve}{\varepsilon}
\newcommand{\twomega}{2\hspace*{.05ex}\omega}
\def\theequation{\thesection.\arabic{equation}}
\begin{document}
\baselineskip 0.8cm
\setcounter{page}{0}
\thispagestyle{empty}
\hfill
\parbox{3cm}{SHEP-93/94-18\\May 1994}
\vspace*{2cm}
\begin{center}
\centerline{\Large Monoslepton production in hadronic collisions}
\vspace*{3cm}
John M$^{\rm c}$Curry\footnote{E-mail: jmcc@hep.phys.soton.ac.uk}\\
\vspace*{0.5cm}
{\it Theory Group, Department of Physics,\\ University of Southampton,
\\ Southampton SO9 5NH, ENGLAND}
\\
\vspace*{2.5cm}
{\large ABSTRACT}
\\
\vspace*{0.5cm}
\parbox{13cm}{\baselineskip 0.8cm
Single sparticle creation in high energy collisions
as a consequence explicit $R$-parity breaking, could be a rich
source of highly spectacular signals at future colliders. One
particular process, which could lead to a highly exotic leptonic
signal at the LHC, is monoslepton production. In this
paper we qualitatively discuss the constraints on the signal
for this process and calculate the hadronic monoslepton
production cross section, taking into account leading QCD
corrections. Our results show the
leading corrections could be quite significant at the proposed
LHC operating energy.}
\end{center}
\newpage
\pagestyle{arabic}
\section{Introduction}
It has been shown\cite{BHR} that certain superstring theories possess natural
discrete symmetries which allow $R$-parity breaking\cite{HS} but prevent
unacceptably fast proton decay by excluding a subset of the possible
$R$-parity breaking operators from the superpotential. Low energy
supersymmetric models naturally arising out of such theories admit
an acceptable phenomenology and thus appear as plausible as the MSSM.
Such $R_p$-broken models therefore deserve to be
confronted with a degree of experimental interest commensurate
with that being directed towards searching for the MSSM.
In minimal $R_p$-broken models the most general allowed renormalizable
superpotential has the form
\be
W = W_{\rm MSSM} + W_{\not{R}_p},
\ee
where $W_{\rm MSSM}$ is the usual mass generating superpotential of the
MSSM, and
\be
W_{\not{R}_p} =
\lambda_{ijk}L_iL_j\overline{e}_k +
\left\{
      \begin{array}{l}
      \lambda_{ijk}^{'}L_iQ_j\overline{d}_k\\
      \hfil \mbox{ XOR} \hfil \\
      \lambda_{ijk}^{''}\overline{u}_i\overline{d}_j\overline{d}_k
      \end{array}
\right.
\label{eq:Rbreak}
\ee
The transformation properties of the matter multiplets under
SU(3)$_{\rm C}\times$SU(2)$_{\rm L}\times$U(1)$_{\rm
Y}$ are listed in Table~\ref{table:trans} below.
\begin{table}[h]
\begin{center}
\begin{tabular}{ c  c}\\ \hline
Superfield  & Quantum Numbers\\ \hline\hline
$L$ & ({\bf 1}, {\bf 2}, {\bf -1} )\\
$\overline{e}$ & ({\bf 1}, {\bf 1}, {\bf 2} )\\
$Q$ & ({\bf 3}, {\bf 2}, {\bf 1/3} )\\
$\overline{u}$ & (\mbox{$\bf \overline{3}$}, {\bf 1}, {\bf -4/3} )\\
$\overline{d}$ & (\mbox{$\bf \overline{3}$}, {\bf 1}, {\bf 2/3} )\\
\hline
\end{tabular}
\caption{Quantum numbers of matter multiplets under
SU(3)$_{\rm C}\times$SU(2)$_{\rm L}\times$U(1)$_{\rm
Y}$ \label{table:trans} }
\end{center}
\end{table}
In eq.~(\ref{eq:Rbreak}) the `XOR' (exclusive or), which is necessary
to avoid extremely fast proton decay, is provided by the aforementioned
stringy discrete symmetries\cite{BHR}. $W_{\not{R}_p}$ generates
new lepton- and baryon-number violating Yukawa interactions which can
lead to spectacular signals in high energy collisions\cite{spectac}.
In this paper we examine the novel possibility of monoslepton
(or anti-slepton) production in hadron collisions induced by one of
the $\left[\,L\,Q\,\overline{d}\,\right]_F$ operators. At tree level
this process proceeds via the graph shown in
Fig.~1, where $i$, $j$ and $k$ are flavour indices.
\begin{center}
\makebox[148pt]{\epsfbox{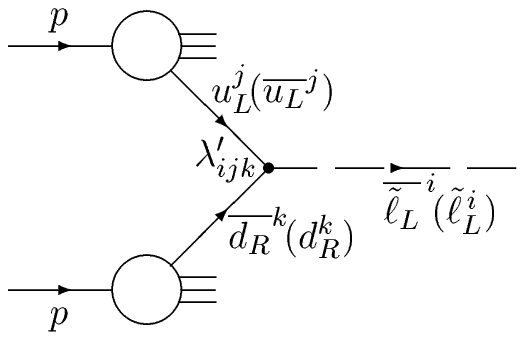}}\\
Figure~1: Single anti-slepton production at leading order.\\
\end{center}
If there is a dominant
$\left[\,L\,Q\,\overline{d}\,\right]_F$
operator the tree level hadronic cross section has the form
\be
\sigma_H^0 = K_0\int_{\tau}^1\frac{{\rm d} x}{x}H(x, \tau/x).
\label{eq:sigmaHtl}
\ee
Here $\tau = M^2/S$, $K_0 = \pi{\lambda^{'}}^2/12S$, where $M$ is the
slepton mass, $\lambda^{\prime}$ is the Yukawa coupling of the
dominant operator, and the hadronic CMS energy is $\sqrt{S}$.
In order to extract any numerical results for the cross section it is
necessary to invoke some assumptions about the flavour structure
of the dominant operator. In this paper we shall assume that
$\lambda^{\prime} = \lambda^{\prime}_{i11}$ ($i = 1,2,3$).
Since the valence up quark content of the proton is greater than the
valence down quark content, this particular flavour structure will favour
anti-slepton production to slepton production. The parton kernel
$H(x_1, x_2)$ for the dominant anti-slepton production mode
is given in terms of the up and anti-down quark
densities ($u$, $\overline{d}$) by
\be
H(x_1,x_2) = u(x_1)\overline{d}(x_2) + u(x_2)\overline{d}(x_1) .
\ee
Assuming B$(\tilde{\ell}\rightarrow \tilde{\gamma} \ell)\sim 1$
and that the photino is the LSP, the slepton/anti-slepton will decay to
produce a dilepton and jets. The
photino, being a Majorana fermion, can couple to both the
$\not{\!\!R}_p$ operators and their charge conjugates which
could allow the production of an unusual
{\em isolated like-sign dilepton} (ILSD). Whether one or both leptons are
isolated depends essentially on the photino to slepton mass ratio:
$\rho = m_{\tilde{\gamma}}/m_{\tilde{\ell}}$. If the photino is very
light ({\em i.e.} $\rho \ll 1$) the secondary lepton could be `lost' in the
hadronic shower and the signal would then appear to be a very
energetic isolated lepton recoiling against a jet. However, assuming that
the secondary lepton can be resolved some qualitative kinematic
constraints on the dilepton are as follows:
\begin{enumerate}
\item
Neglecting lepton masses, the energy of the primary lepton ($l_1$) is
\be
E_1 =
\frac{m_{\tilde{\ell}}}{2}\left(1-\rho^2\right),
 \ee
and $E_2\in \left[\mbox{max}(E_1 - m_{\tilde{\gamma}}/2,0),
E_1 + m_{\tilde{\gamma}}/2\right]$ for the secondary lepton ($l_2$).
\item
For a moderately light photino ({\em i.e.}
$\rho \leq (\sqrt{5}-1)/2 \approx 0.62$) the minimum kinematically allowed
opening angle between the leptons is
\be
\theta_{min}(\rho) = \pi -\sin^{-1}\left(\frac{\rho}{1-\rho^2}\right)
\:\in \left[\frac{\pi}{2}, \pi \right]
\ee
If $\rho > 0.62$ there is no constraint.
\item
The missing transverse momentum carried by the leptons can be as large as
$m_{\tilde{\gamma}}/2$.
\end{enumerate}
Potentially large backgrounds from $t \rightarrow b\,W^+$ can be
effectively excluded with a suitable cut on
$E_1$ ({\em e.g.} $E_1 > O(m_t/2)$),
whilst $q\overline{q},\,H^0 \rightarrow W^+W^-$
can be eliminated on the basis of event topology (if the secondary
lepton is isolated), or if necessary with $M_{\rm jets} \neq
O(M_W)$. A further cut $M_{{l_1}{l_2}} \neq O(M_{Z^0})$
may be necessary to veto `fake' energetic $l^+l^-$ pairs from high-$p_T$ $Z^0$
production giving an ILSD signal. In any case, after the standard
model is removed monoslepton production at the
LHC will have an exotic signature.
\par
Turning our attention to the cross section we will now show
that the leading QCD corrections to the
tree level expression in eq.~(\ref{eq:sigmaHtl}) are potentially
large at proposed LHC energies (in fact $>30\%$ for
$M \stackrel{>}{_{\sim}}1$TeV). Our
calculation of these corrections closely follows the approach
adopted by AEM (Altarelli, Ellis and Martinelli \cite{AEM})
in their calculation of the leading QCD corrections to vector boson
production. We, of course, extend the AEM calculation
to consider additional one loop vertex and self energy corrections involving
virtual squarks and gluinos.

\setcounter{equation}{0}
\section{The NLO corrections}
\subsection{Overview of calculation}
The treatment of the NLO corrections to the total cross section
is a straightforward perturbative calculation. The full
$O(\alpha_s)$ contribution is obtained by adding the interference terms
generated by the tree level diagram in Fig.~2
with the one loop diagrams in Fig.~3(a)-(d) to
the sum of the `squares' of the tree level diagrams in Fig.~4(a)-(d).
\begin{center}
\makebox[115pt]{\epsfbox{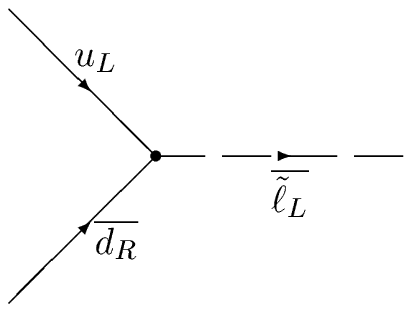}}\\
Figure~2:~The leading order contribution to the elementary cross
section\\
\end{center}
\par
We evaluate the contribution from these diagrams as follows: Using
dimensional regularisation to tackle both the UV and IR
singularities we initially continue to
$4-2\ve$ dimensions ($\ve > 0$) to address the UV divergences.
Since IR divergences are manifest in $4-2\ve$ dimensions we
artificially `postpone' the mass-singularity in the quark self energy
by temporarily continuing external-line quarks off-shell. After
renormalization of the UV divergences we then continue to
$4+2\eta$ dimensions ($\eta > 0$) and return all external
quarks on-shell. The IR divergences are now regularised and
appear as single and double poles in $\eta$. The double poles, which
originate from vanishingly soft virtual gluon emission/absorption
processes, cancel when the contributions from the gluon bremsstrahlung and
the one loop QCD diagrams are added. The single pole
terms do not cancel completely, but those which survive
({\em i.e.} the `mass singularities') are  removed
in the standard manner\cite{AEM} by factoring them into
the `bare' NLO parton densities. We shall now briefly consider the
contribution from the individual diagrams, starting with the loop
diagrams.
\begin{center}
\makebox[293pt]{\epsfbox{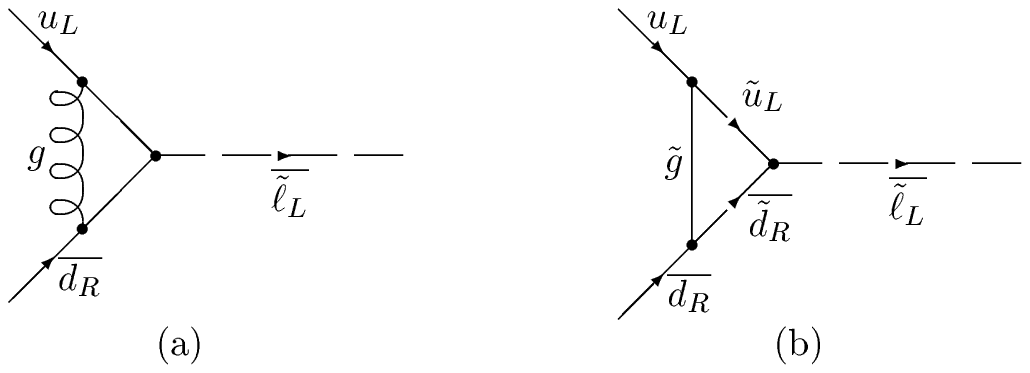}}\\
Figure~3(a),(b): The one-loop QCD and SSB vertex corrections\\
\makebox[291pt]{\epsfbox{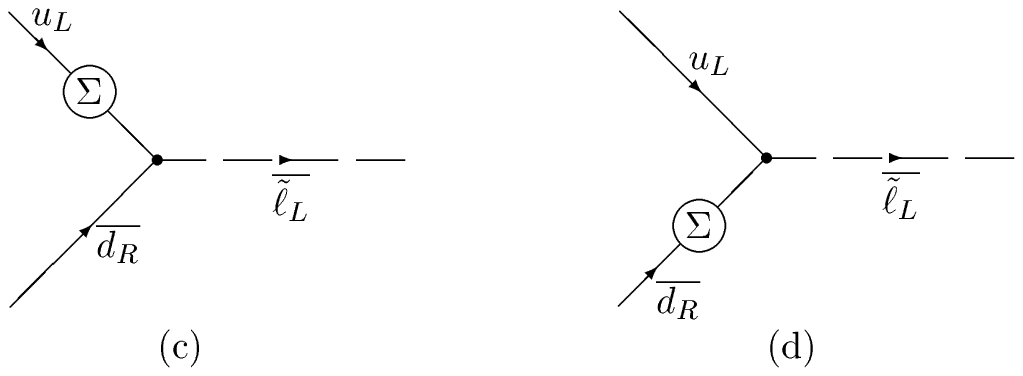}}\\
Figure~3(c),(d): The quark self-energy corrections
($\Sigma=\Sigma_{\rm QCD}+\Sigma_{\rm SUSY}$) \\
\end{center}
\begin{center}
\makebox[301pt]{\epsfbox{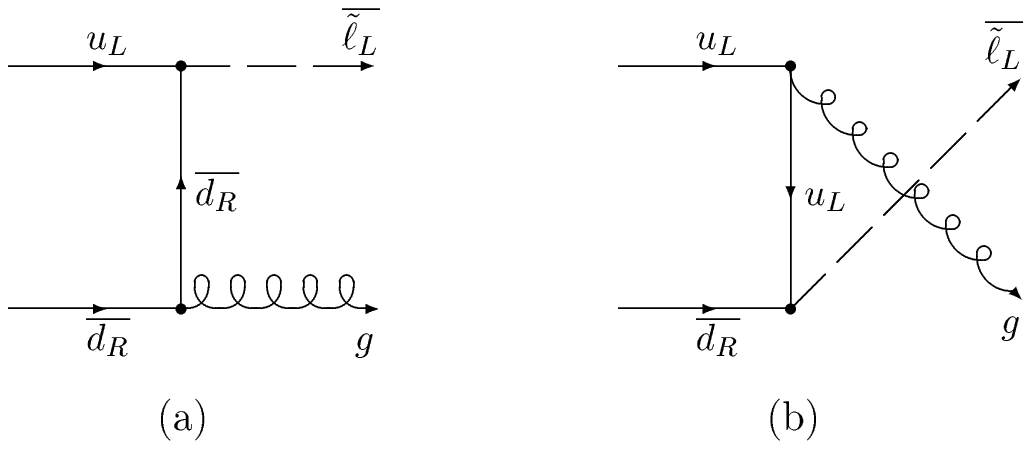}}\\
Figure~4(a),(b): The NLO `gluon bremsstrahlung' corrections\\
\end{center}
\begin{center}
\makebox[298pt]{\epsfbox{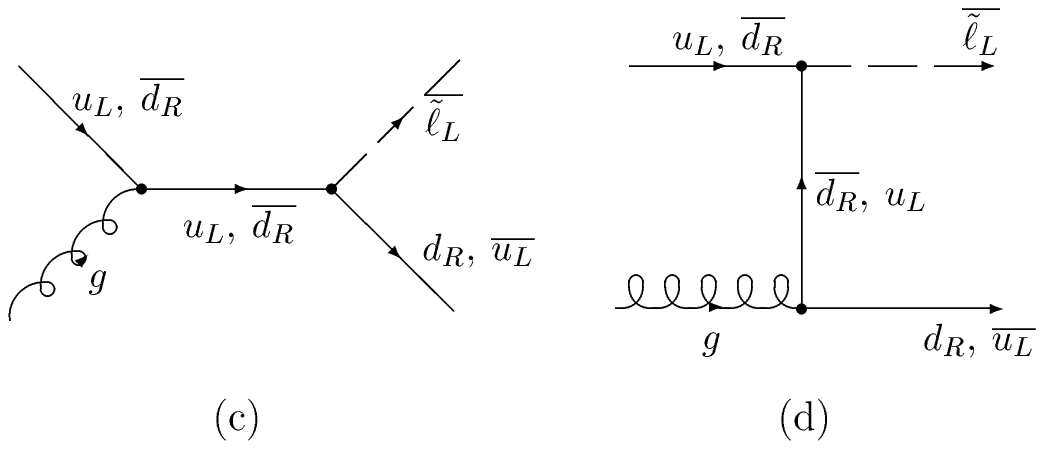}}\\
Figure~4(c),(d): The NLO `Compton scattering' corrections\\
\end{center}

\vspace*{1cm}
\subsection{The one-loop contribution}

The net effect of all of the loop diagrams (evaluated in
$4-2\ve$ dimensions) is to simply multiply the
tree level cross section by a factor
\be
1+\frac{\alpha_s\,C_F}{\pi}L\left(\frac{1}{\ve^\prime}
;\frac{1}{\ve},\frac{1}{\ve^2}\right),
\label{eq:Loopfactor}
\ee
where $\alpha_s$ is the dimensionless strong coupling, $C_F$ is the
quadratic Casimir in the fundamental representation
of the gauge group ($C_F = 4/3$  for SU(3)), and $L$ is some
function to be determined. Formally, $\ve^{\prime} = \ve$ but the
prime is used to indicate that the simple pole is of UV and not
IR origin. Retaining only $O(\alpha_s)$ terms it is clear that we
can re-write the multiplicative factor of~(\ref{eq:Loopfactor})
in the convenient factorised form
\be
\left(
1+\frac{\alpha_s\,C_F}{\pi}L_{\rm UV}\left(\frac{1}{\ve^\prime}\right)
\right)
\left(
1+\frac{\alpha_s\,C_F}{\pi}L_{\rm IR}
\left(\frac{1}{\ve},\frac{1}{\ve^2}\right)
\right)
\label{eq:loopfactors}
\ee
where $L_{\rm UV}$ and $L_{\rm IR}$ contain the UV and IR poles
respectively. Unfortunately, finite corrections ensure
that $L_{\rm UV}$ and $L_{\rm IR}$ cannot be uniquely determined.
\\[3ex]
{\bf
The QCD vertex correction\\[2ex]}
The overall effect of including the QCD vertex correction of
Fig.~3(a) (calculated in the Feynman gauge) is to multiply the
tree level cross section by the factor
\be
1+\frac{\alpha_{s}C_F}{2\pi}
\left(\frac{ {\bar{\mu}}^{2} }{M^{2}}\right)^{\ve}
\left(\frac{-2}{\ve^{2}}+\frac{4}{\ve^{\prime}}-
\frac{4}{\ve} + \frac{7\pi^{2}}{6} -2 + O(\ve)\right).
\label{eq:vfac1}
\ee
Here $M$ is the slepton mass, $\bar{\mu} =4\pi\mu\,e^{-\gamma_E}$
where $\gamma_{E}$ is the Euler-Mascheroni constant,
and $\mu$ is an arbitrary scale introduced to ensure that $\alpha_s$
is always dimensionless:
\be
\alpha_{s} = \frac{(\mu^{2})^{-\ve}g^{2}}{4\pi}.
\ee
For future convenience we will factor out the UV pole piece and
re-write~(\ref{eq:vfac1}) in the `UV$\times$IR' form
(neglecting $O(\alpha_{s}^{2},\ve)$):-
\be
\left[1+\frac{\alpha_{s}C_F}{2\pi}
\left(\frac{ {\bar{\mu}}^{2} }{M^{2}}\right)^{\ve}
\frac{4}{\ve^{\prime}}\right]
\left[1+\frac{\alpha_{s}C_F}{2\pi}
\left(\frac{ {\bar{\mu}}^{2} }{M^{2}}\right)^{\ve}
\left(\frac{-2}{\ve^{2}}-
\frac{4}{\ve} + \frac{7\pi^{2}}{6} -2\right)\right].
\label{eq:finalqcdvert}
\ee
\\[3ex]
{\bf
The soft supersymmetry breaking vertex correction\\[2ex]}
The soft supersymmetry breaking (SSB) vertex correction of diagram
Fig.~3(b) possesses neither UV or IR singularities, but it does
introduce a dependence on squark and gluino masses. In fact, the effect
of the this  diagram is to multiply the tree level cross section by
the finite factor
\be
1+\frac{\alpha_{s}C_F}{2\pi}\cdot
\frac{4\,A\,\xi_{ \tilde{g} } }{\lambda^{\prime}}\cdot
\Delta_{\rm V}(\xi_{\tilde{u}}^{2},\xi_{\tilde{d}}^{2},\xi_{\tilde{g}}^{2})
\ee
where, $A$ is the {\em dimensionful}\, soft breaking tri-scalar coupling,
$\xi_{\tilde{u}} = m_{\tilde{u}}/M$, $\xi_{\tilde{d}} = m_{\tilde{d}}/M$
and $\xi_{\tilde{g}} = m_{\tilde{g}}/M$. The factor
$\Delta_{\rm V}(\xi_{\tilde{u}}^{2},
\xi_{\tilde{d}}^{2},\xi_{\tilde{g}}^{2})$ is given by
\be
\Delta_{V}(\xi_{\tilde{u}}^{2},\xi_{\tilde{d}}^{2},\xi_{\tilde{g}}^{2})
= \int_{0}^{1}{\rm d} x \,\frac{1}{\mu_{+}(x)+\mu_{-}(x)}\,\ln\left[
\frac{\mu_{-}(x)^{2}-(\xi_{\tilde{u}}^2-\xi_{\tilde{d}}^2)^2}
{\mu_{+}(x)^{2}-(\xi_{\tilde{u}}^2-\xi_{\tilde{d}}^2)^2}
\right],
\ee
where
\be
\mu_{\pm}(x) = \left[
x^2 + 2(2\xi_{\tilde{g}}^2-\xi_{\tilde{u}}^2-\xi_{\tilde{d}}^2)x +
(\xi_{\tilde{u}}^2-\xi_{\tilde{d}}^2)^2 -
4\xi_{\tilde{g}}^2 \right]^{\frac{1}{2}} \pm x.
\ee
Of course, $\Delta_{V}(\xi_{\tilde{u}}^{2},\xi_{\tilde{d}}^{2},
\xi_{\tilde{g}}^{2})$ may be calculated numerically for any given
$\xi_{\tilde{u}}^2$, $\xi_{\tilde{d}}^2$ and
$\xi_{\tilde{g}}^2$, but we note that for the case in which the
sparticles are degenerate it can actually be performed
analytically to yield a value of $\pi^2/72$.
\\[3ex]
{\bf
The QCD contribution to the quark self-energy\\[2ex]}
To calculate the contribution from the self energy correction we
temporarily shift the external quarks off-shell which enables
us to write down a well defined Feynman amplitude for
diagrams (c) and (d) in Fig.~3. The one loop self energy
kernel `$\Sigma(p)$' has, in addition to
the standard QCD contribution, a supersymmetric contribution involving
squarks and gluinos. In obvious notation we shall therefore
write $\Sigma(p) = \Sigma_{\rm QCD}(p) + \Sigma_{\rm SUSY}(p)$.
By Lorentz covariance $\Sigma(p)$ must have an expansion in momentum
of the form
\be
\Sigma(p) = A(p^{2})+B(p^{2})\not{\!p}.
\ee
One can easily show that the net effect of including the self energy
(SE) correction is to multiply the tree level cross section by the
factor
\be
1+2\,{\cal R}e \,B(0),
\label{eq:secont}
\ee
where $B(0)$ is (naively) obtained from the self-energy by
\be
B(0) = \lim_{p^{2}\rightarrow 0}
\,\frac{\partial\,\Sigma(p)}
{\partial\!\not{\!p}}.
\label{eq:B0}
\ee
The limit in eq.~(\ref{eq:B0}) must be taken with care since
$\Sigma_{\rm QCD}(p)$ is singular at $p^{2} = 0$
when calculated in less than four dimensions. We shall therefore keep
$p^2 \neq 0$ until the UV divergences have been renormalized and we
have continued to $4+2\eta$ dimensions to address the IR
divergences. Like the QCD vertex correction $\Sigma_{\rm QCD}(p)$
is gauge dependent and possesses both UV and IR singularities.
The Feynman gauge contribution to the SE from Fig.~5 is
\be
\Sigma_{\rm QCD}(p) = -\frac{\alpha_{s}C_F}{4\pi}\,\not{\!p}\,
\left(\frac{{\bar{\mu}}^{2}}{-p^{2}}\right)^{\ve}
\left(\frac{1}{\ve^{\prime}}+1+O(\ve)\right),
\label{eq:QCDSE1}
\ee
where the prime on $\ve$ indicates a UV pole. Thus, from
eq.~(\ref{eq:secont}) we can see that the effect of the QCD SE correction
is to multiply the tree level cross section by the factor
\be
1-\frac{\alpha_{s}C_F}{2\pi}\,
\left(\frac{{\bar{\mu}}^{2}}{-p^{2}}\right)^{\ve}
\left(\frac{1}{\ve^{\prime}}+1\right)
\label{eq:sefac1}
\ee
which, up to $O(\alpha_{s}^{2},\ve)$ terms, can be written in
the desired UV$\times$IR form
\be
\left[
1-\frac{\alpha_{s}C_F}{2\pi}\,
\left(\frac{{\bar{\mu}}^{2}}{M^{2}}\right)^{\ve}
\frac{1}{\ve^{\prime}}
\right]
\left[
1+\frac{\alpha_{s}C_F}{2\pi}\,
\left(\frac{{\bar{\mu}}^{2}}{M^{2}}\right)^{\ve}
\left(\frac{1}{\ve}-\left(\frac{M^{2}}{-p^{2}}\right)^{\ve}
\left(1+\frac{1}{\ve}\right)
\right)
\right].
\ee
\begin{center}
\hspace*{-1cm}
\makebox[136pt]{\epsfbox{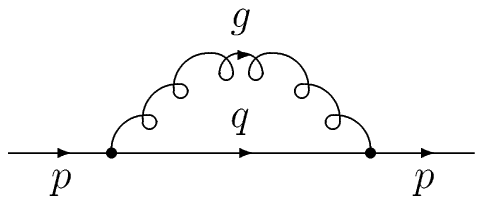}}\\
Figure~5:~The QCD contribution to the quark self-energy\\[3ex]
\end{center}
{\bf
The supersymmetric contribution to the quark self-energy\\[2ex]}
The contribution to the self-energy from diagram Fig.~6 is
\be
\Sigma_{\rm SUSY}(p) = -\frac{\alpha_{s}C_F}{4\pi}\,\not{\!p}\,\left[
\left(\frac{{\bar{\mu}}^{2}}{M^{2}}\right)^{\ve}
\frac{1}{\ve^{\prime}}- 2\int_{0}^{1} {\rm d} x \,x\,\ln
\left(x(\xi_{\tilde{q}}^{2}-
\xi_{\tilde{g}}^{2})+
\xi_{\tilde{g}}^{2}\right)\right],
\label{eq:seintegral}
\ee
where $\tilde{q}$ is some generic squark. From eq.~(\ref{eq:secont}) we
can deduce that the effect of the supersymmetric self-energy
correction is to simply multiply the tree level cross section by the factor :-
\be
1+\frac{\alpha_{s}C_F}{2\pi}\left[
\Delta_{\rm SE}(\xi_{\tilde{u}}^{2},\xi_{\tilde{d}}^{2},\xi_{\tilde{g}}^{2})
-\left(\frac{{\bar{\mu}}^{2}}{M^{2}}\right)^{\ve}
\frac{1}{\ve^{\prime}}
\right]
\label{eq:ssselast}
\ee
where the sparticle mass dependent term `$\Delta_{\rm SE}(\xi_{\tilde{u}}^{2},
\xi_{\tilde{d}}^{2},\xi_{\tilde{g}}^{2})$' is given by
\be
\Delta_{\rm SE}(\xi_{\tilde{u}}^{2},\xi_{\tilde{d}}^{2},\xi_{\tilde{g}}^{2})
 = \int_{0}^{1} {\rm d} x \,x\left[
\ln\left(x(\xi_{\tilde{u}}^{2}-
\xi_{\tilde{g}}^{2})+
\xi_{\tilde{g}}^{2}\right)
+\ln\left(x(\xi_{\tilde{d}}^{2}-
\xi_{\tilde{g}}^{2})+
\xi_{\tilde{g}}^{2}\right)
\right],
\ee
which can be performed analytically. For a degenerate sparticle mass
spectrum we have the trivial result $\Delta_{\rm SE} = 0$, whilst for
the non-degenerate case we find
\begin{eqnarray}
\Delta_{\rm
SE}(\xi_{\tilde{u}}^{2},\xi_{\tilde{d}}^{2},\xi_{\tilde{g}}^{2})
&=& \ln(\xi_{\tilde{u}}\xi_{\tilde{d}}) -
\frac{ \ln(\xi_{\tilde{u}}/\xi_{\tilde{g}}) }
{( 1-\xi_{\tilde{u}}^2 / \xi_{\tilde{g}}^2)^2 }
-\frac{ \ln(\xi_{\tilde{d}}/\xi_{\tilde{g}}) }
{( 1-\xi_{\tilde{d}}^2 / \xi_{\tilde{g}}^2)^2 }
\nonumber\\
&&-\frac{1}{2}
\left[1+
\frac{1}{ 1-\xi_{\tilde{u}}^2 / \xi_{\tilde{g}}^2 }
+\frac{1}{ 1-\xi_{\tilde{d}}^2 / \xi_{\tilde{g}}^2 }
\right].
\label{eq:deltase}
\end{eqnarray}
Finally, factoring out the UV pole of~(\ref{eq:ssselast})
in the usual manner we find that the multiplicative factor
produced by the supersymmetric self-energy correction is
\be
\left[
1-\frac{\alpha_{s}C_F}{2\pi}\left(\frac{{\bar{\mu}}^{2}}
{M^{2}}\right)^{\ve}\frac{1}{\ve^{\prime}}
\right]
\left[
1+\frac{\alpha_{s}C_F}{2\pi}\Delta_{\rm SE}(\xi_{\tilde{u}}^{2},
\xi_{\tilde{d}}^{2},\xi_{\tilde{g}}^{2})
\right]
\label{eq:sefactor}
\ee
Having calculated all the one loop contributions we shall now remove
the UV singularities before considering the tree level corrections.
\begin{center}
\hspace*{-1cm}
\makebox[143pt]{\epsfbox{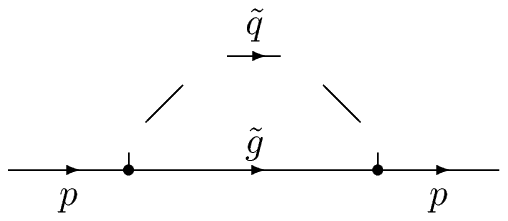}}\\
Figure~6:~The supersymmetric contribution to the quark self-energy\\
\end{center}

\subsection{Renormalization}
It is convenient to make the
following choices for $L_{\rm UV}$ and  $L_{\rm IR}$ :-
\begin{eqnarray}
L_{\rm UV}\left(\frac{1}{\ve^{\prime}}\right)
&=&
\left(\frac{{\bar{\mu}}^{2}}
{M^{2}}\right)^{\ve}\frac{1}{\ve^{\prime}}+
\frac{1}{2}
\Delta_{\rm SE}(\xi_{\tilde{u}}^{2},
\xi_{\tilde{d}}^{2},\xi_{\tilde{g}}^{2})
+\frac{2\,A\,\xi_{ \tilde{g} } }{\lambda^{\prime}}
\Delta_{\rm V}(\xi_{\tilde{u}}^{2},\xi_{\tilde{d}}^{2},\xi_{\tilde{g}}^{2}),
\label{eq:Luv}
\\
L_{\rm IR}\left(\frac{1}{\ve},\frac{1}{\ve^2}\right)
&=&
\left(\frac{ {\bar{\mu}}^{2} }{M^{2}}\right)^{\ve}
\left[\frac{-1}{\ve^{2}}-
\frac{3}{2\ve} + \frac{7\pi^{2}}{12}
-\frac{1}{2}\left(\frac{M^{2}}{-p^{2}}\right)^{\ve}
\left(1+\frac{1}{\ve}\right)-1
\right].
\label{eq:Lir}
\end{eqnarray}
That is, we have chosen to include the finite corrections from the
soft SUSY breaking graphs in the UV contribution. To address the
problem of renormalization we must identify the
renormalization constants $Z_1,\;Z_2$ and $Z_3$
which are related to each other by
\be
Z_1Z_2 = Z_3
\label{eq:renormrel}
\ee
and are associated with Yukawa coupling, quark wave-function and
Yukawa vertex renormalization respectively. In the $\overline{\rm MS}$
scheme we find
\be
Z_1 = 1-\frac{\alpha_{s}C_F}{2\pi}\frac{1}{\hat{\ve}},\hspace{3ex}
Z_2 = 1-\frac{\alpha_{s}C_F}{2\pi}\frac{1}{\hat{\ve}},\hspace{3ex}
Z_3 = 1-\frac{\alpha_{s}C_F}{\pi}\frac{1}{\hat{\ve}},
\label{eq:Z1}
\ee
where
\be
\frac{1}{\hat{\ve}} =
\frac{1}{\ve}+\ln4\pi-\gamma_{E}.
\ee
As one would expect the renormalization constants $Z_2$
and $Z_3$ are gauge dependent but $Z_1$, pertaining to a physical
observable, is gauge invariant. Calculation of the QCD
loop diagrams in Landau gauge (see Appendix) leads us to conclude
$Z_2$ and $Z_3$ to be
\beqarr
Z_2 = 1-\frac{\alpha_{s}C_F}{4\pi}\frac{1}{\hat{\ve}},&&
Z_3 = 1-\frac{\alpha_{s}C_F}{4\pi}\frac{3}{\hat{\ve}}.
\eeqarr
Thus, we can deduce that in a general $R_{\xi}$ gauge
\beqarr
Z_2 = 1-\frac{\alpha_{s}C_F}{\pi}\left(\frac{\xi+1}{4}\right)
\frac{1}{\hat{\ve}},&&
Z_3 = 1-\frac{\alpha_{s}C_F}{\pi}\left(\frac{\xi+3}{4}\right)
\frac{1}{\hat{\ve}}.
\eeqarr
from which one can see $Z_1$ is still given by the gauge
invariant expression in eq.~(\ref{eq:Z1}). Having determined $Z_1$ one
finds that the $O(\alpha_s)$ beta function for the $R_p$-breaking coupling
$\lambda^{\prime}$ is
\be
\beta_{\lambda^{\prime}} = \mu\,\frac{{\rm d}
\lambda^{\prime}}{{\rm d} \mu} =
-\lambda^{\prime}\frac{\alpha_{s}C_F}{\pi},
\label{eq:rge1}
\ee
which clearly shows that $\lambda^{\prime}$
decreases as the scale $\mu$ is increased.
Solving eq.~(\ref{eq:rge1})
we find
\be
\lambda^{\prime}(M^{2}) = \lambda^{\prime}(\mu^{2})
\left[1+\frac{\alpha_{s}C_F}{2\pi}
\ln\left(\frac{\mu^{2}}{M^{2}}\right)\right]+O(\alpha_{s}^{2}).
\label{eq:rgesoln}
\ee
Before moving on to consider the resultant renormalized contribution to the
cross section, we note that by performing an additional finite
renormalization on $\lambda^{\prime}$ we can `remove' the residual
finite terms in eq.~(\ref{eq:Luv}). In all subsequent work we shall
assume that this finite renormalization has been performed.
\par
Adding the contributions from the renormalized loop graphs to the
leading tree level graph we find the resultant to be
\be
K_{0}\,(M^{2})\,
\left(
1+\frac{\alpha_s\,C_F}{\pi}L_{\rm IR}
\left(\frac{1}{\ve},\frac{1}{\ve^2}\right)
\right)
\ee
where the `running' tree level cross section is
\be
K_{0}\,(M^{2}) = \frac{2^{-\ve}\pi{\lambda^{\prime}}^{2}(M^{2})}{12S}.
\ee
Thus, the net effect of the UV renormalization is to
simply replace the scale independent Yukawa coupling
$\lambda^{\prime}$ by a running
coupling $\lambda^{\prime}(M^2)$ whose evolution is governed by
eq.~(\ref{eq:rgesoln}).
Since the UV renormalization procedure is now complete, we can now
continue from $4-2\ve$ dimensions to $4+2\eta$ (with $\eta>0$)
dimensions to regularise the IR divergences. In $4+2\eta$ dimensions we
can safely take the $p^{2}\rightarrow 0$ limit
of $L_{\rm IR}$ in eq.~(\ref{eq:Lir}) and obtain the well defined result
\be
L_{\rm IR}\left(\frac{1}{\eta},\frac{1}{\eta^2}\right)
= \left(\frac{M^{2}}{ \bar{\mu}^{2} }\right)^{\eta}
\left[-\frac{1}{\eta^{2}}+\frac{3}{2\eta}
+\frac{7\pi^{2}}{12} -1\right].
\ee

\subsection{The NLO tree level contribution}
The two body phase space (in $\twomega$ dimensions)
for the on-shell particle pair production diagrams in Fig.~4 is
({\em c.f.} \cite{AEM})
\be
{\rm d}\Phi_{(2)} = \frac{{\rm d}^{\twomega} p_{3}}{(2\pi)^{\twomega}}
		    \frac{{\rm d}^{\twomega} p_{4}}{(2\pi)^{\twomega}}
		  \cdot (2\pi)^{\twomega} \delta(p_{1}+p_{2}-p_{3}-p_{4})
		  \cdot 2\pi \delta^{+}(p_{3}^{2}-M^{2})
		  \cdot 2\pi \delta^{+}(p_{4}^{2}),
\ee
where $(p_1,p_2)$ and $(p_3,p_4)$ denote the momenta of the incoming
and outgoing particles respectively, with $p_3$ being the momentum of the
slepton. The partonic Mandlestam variables ($s,t,u$) are defined in the
usual way
by
\beqarr
s = (p_1+p_2)^2, & t = (p_1-p_3)^2,& u = (p_1-p_4)^2,
\eeqarr
and in terms of $w = \frac{1}{2}
(1+\hat{\mbox{\boldmath $p$}}_{4}\cdot\hat{\mbox{\boldmath $p$}}_{2})$
and $z = M^2/s$ we find $t$ and $u$ are
\be
t = -\frac{M^{2}}{z}(1-z)(1-w),\hspace*{4ex}u = -\frac{M^{2}}{z}(1-z)w.
\label{eq:utinwz}
\ee
Using these identities to write the differential phase space in terms of
$z$ and $w$ we find after some simple algebra
\be
\frac{{\rm d}\Phi_{(2)}} {\!{\rm d}w\:} =
\frac{1}{8\pi\Gamma(\omega-1)}
\left(\frac{M^{2}}{4\pi}\right)^{\omega-2}\!(1-z)
\left[\frac{(1-z)^{2}}{z}\right]^{\omega-2}(w(1-w))^{\omega-2}.
\label{eq:PS}
\ee
Setting $\eta = \omega-2$ and using the result\cite{AandS}
\be
\Gamma(1+\xi) = \exp\left(-\gamma_{E}\xi+\frac{\pi^{2}}{12}\xi^{2}\right)
+O(\xi^{3})
\label{eq:Gammaexpansion}
\ee
we find eq.~(\ref{eq:PS}) becomes
\be
\frac{{\rm d}\Phi_{(2)}} {\!{\rm d}w\:} =
\frac{(\mu^{2})^{\eta}}{8\pi}
\left(\frac{M^{2}}{{\bar{\mu}}^{2}}\right)^{\eta}
\frac{(1-z)^{1+2\eta}}{z^{\eta}}\,
\left(1-\frac{\pi^{2}}{12}\eta^{2}\right)
\,(w(1-w))^{\eta}+O(\eta^{3}),
\label{eq:diffangps}
\ee
where we have explicitly introduced an arbitrary mass scale `$\mu$'.
Thus, contribution to the elementary cross section (in $4+2\eta$
dimensions) from any one of the diagrams in Fig.~4 has the form
\be
\hat{\sigma} = \frac{(\mu^{2})^{\eta}}{16\pi\,s}
\left(\frac{M^{2}}{{\bar{\mu}}^{2}}\right)^{\eta}
\frac{(1-z)^{1+2\eta}}{z^{\eta}}\,
\left(1-\frac{\pi^{2}}{12}\eta^{2}\right)
\int_{0}^{1}{\rm d}w\,
\,(w(1-w))^{\eta}\,\langle\,\left|{\cal M}(z,w)\right|^{2}\rangle_{(\eta)}
+O(\eta),
\label{eq:sigmahat}
\ee
where $\langle\,\left|{\cal M}(z,w)\right|^{2}\rangle_{(\eta)}$ denotes
the appropriate (spin/colour averaged) squared Feynman amplitude
in $4+2\eta$ dimensions.

Turning our attention to the `gluon-bremsstrahlung' diagrams in
Fig.~4(a),~(b) we find after a brief calculation
\be
\langle\, \left| {\cal M}_{a}+{\cal M}_{b} \right| ^{2}\rangle =
16\pi\,S\,K_{0}\,(M^{2})\,
\frac{g^{2}C_F}{8\pi^{2}}\,
\left[\frac{(s-M^{2})^{2}}{ut}(1+\eta)+\frac{2sM^{2}}{ut}\right].
\label{eq:maandmb3}
\ee
In terms of $w$ and $z$ eq.~(\ref{eq:maandmb3}) becomes
\be
\langle\, \left| {\cal M}_{a}+{\cal M}_{b} \right| ^{2}\rangle =
\frac{16\pi\,S\,K_{0}\,(M^{2})}{w(1-w)}\,
\frac{g^{2}C_F}{8\pi^{2}}\,
\left[\frac{1+z^{2}}{(1-z)^{2}}+\eta\right],
\ee
and so from eq.~(\ref{eq:sigmahat}) we find that the
contribution of the gluon bremsstrahlung diagrams
to the elementary cross section is
\be
K_{0}\,(M^{2})\,\frac{S}{s}\,
\frac{\alpha_{s}C_F}{2\pi}\,
\left(\frac{M^{2}}{{\bar{\mu}}^{2}}\right)^{\eta}
\frac{(1-z)^{1+2\eta}}{z^{\eta}}\,
\left(1-\frac{\pi^{2}}{12}\eta^{2}\right)
\left[\frac{1+z^{2}}{(1-z)^{2}}+\eta\right]\,I(\eta)+O(\eta),
\label{eq:beforeint}
\ee
where
\be
I(\eta) = \int_{0}^{1}{\rm d}w\,
\,(w(1-w))^{\eta -1}.
\ee
This integral can calculated directly in terms of the Euler Beta function:
\be
I(\eta) = B(\eta,\eta) =
\frac{2}{\eta}-\frac{\pi^{2}}{3}\eta
+O(\eta^{2}),
\ee
where we have used eq.~(\ref{eq:Gammaexpansion}) to obtain the power
series expansion. With this result~(\ref{eq:beforeint}) becomes
\be
K_{0}\,(M^{2})\,\frac{S}{s}\,
\frac{\alpha_{s}C_F}{\pi}\,
\left(\frac{M^{2}}{{\bar{\mu}}^{2}}\right)^{\eta}
\left(1-\frac{\pi^{2}}{4}\eta^{2}\right)
\left[\frac{z^{-\eta}}{\eta}(1+z^{2})(1-z)^{-1+2\eta}
+z^{-\eta}(1-z)^{1+2\eta}\right]
+O(\eta).
\label{eq:sigmab1}
\ee
After isolating the soft gluon pole at $z = 1$ and performing a power
series expansion in $\eta$, we find the content of the square bracket in
eq.~(\ref{eq:sigmab1}) is
\be
\frac{1}{\eta^{2}}\,\delta(1-z) + \frac{1}{\eta}\,\frac{1+z^{2}}{(1-z)_{+}} +
(1+z^{2})\,\left(\frac{\ln (1-z)^{2}}{1-z}\right)_{+} -
(1+z^{2})\,\frac{\ln z}{1-z} + 1-z + O(\eta),
\ee
where the `+' indicates conventional distributional
regularisation on $[0,1]$. Upon addition of the contributions from
the gluon-bremsstrahlung diagrams
and the loop diagrams (which have support only at $z=1$) the double
poles cancel and the resulting contribution `$\Delta\,\hat{\sigma}$'
to the elementary cross section is
\be
\Delta\,\hat{\sigma} =
K_{0}\,(M^{2})\,
\frac{S}{s}\,\left[\delta(1-z)+\frac{\alpha_{s}C_F}{2\pi}
\left(
2\,P_{qq}(z)\ln\left(\frac{M^{2}}{\mu^{2}}\right)
+ A(z)
\right)
\right]+O(\alpha_{s}^{2}),
\label{eq:quarkbit}
\ee
where
\be
P_{qq}(z) = \frac{1+z^{2}}{(1-z)_{+}}+\frac{3}{2}\,\delta(1-z),
\ee
is the quark $\rightarrow$ quark `splitting function',  and
\begin{eqnarray}
A(z) &=& 2\,\left(\frac{1}{\eta}-\ln4\pi+\gamma_{E}\right)P_{qq}(z) +
\left(\frac{2\pi^{2}}{3}-2\right)\,\delta(1-z) + 2\,(1-z)\nonumber\\
&&+ 2\,(1+z^{2})\,\left(\frac{\ln (1-z)^{2}}{1-z}\right)_{+} -
2\,(1+z^{2})\,\frac{\ln z}{1-z}.
\end{eqnarray}
\par
We now turn our attention to the `Compton' like graphs in
Fig.~4(c),(d). The contribution from these graphs is obtained from
eq.~(\ref{eq:maandmb3}) by making the replacement
$s \leftrightarrow - u$. Thus,
\be
\langle\, \left| {\cal M}_{c}+{\cal M}_{d} \right| ^{2}\rangle =
-16\pi\,S\,K_{0}\,(M^{2})\,
\frac{g^{2}T_F}{8\pi^{2}}\,
\left[\frac{(u-M^{2})^{2}}{st}(1+\eta)+\frac{2uM^{2}}{st}\right],
\label{eq:mcandmd2}
\ee
where
\be
T_F\delta^{ab} = {\rm tr}(T^{a}T^{b}) = \frac{1}{2}\delta^{ab}
\ee
is the quadratic trace in the fundamental representation.
In terms of $w$ and $z$ the RHS of eq.~(\ref{eq:mcandmd2}) becomes
\be
16\pi\,S\,K_{0}\,(M^{2})\,
\frac{g^{2}T_F}{8\pi^{2}}\,
\left[\frac{z^{2}}{(1-w)(1-z)}+(1-z)\,\frac{w^{2}}{(1-w)}
+\eta\,\frac{[z+(1-z)w]^{2}}{(1-w)(1-z)}\right].
\ee
Thus, the contribution to the elementary cross section is
\be
K_{0}\,(M^{2})\,\frac{S}{s}\,
\frac{\alpha_{s}T_F}{2\pi}\,
\left(\frac{M^{2}}{{\bar{\mu}}^{2}}\right)^{\eta}
\frac{(1-z)^{1+2\eta}}{z^{\eta}}
\left(1-\frac{\pi^{2}}{12}\eta^{2}\right)
I(z,\eta),
\label{eq:compton}
\ee
where
\beqarr
I(z,\eta) &=&
\int_{0}^{1}{\rm d}w\,
\,(w(1-w))^{\eta}\,
\left[\frac{z^{2}}{(1-w)(1-z)}
+(1-z)\,\frac{w^{2}}{(1-w)}
\right.\nonumber\\
&&\left.+\eta\,\frac{[z+(1-z)w]^{2}}{(1-w)(1-z)}\right].
\eeqarr
Using eq.~(\ref{eq:Gammaexpansion}) to expand the Euler Beta functions
which arise from the above integrals we obtain
\be
I(z,\eta) =
\frac{z^{2}}{1-z}\,\left(\frac{1}{\eta}+1\right) +
(1-z)\,\left(\frac{1}{\eta}-\frac{1}{2}\right) + 2\,z
+ O(\eta).
\ee
Expanding the remaining terms of~(\ref{eq:compton}) as a power series
in $\eta$ we find that the $O(\alpha_s)$ contribution of the Compton
type diagrams to the elementary cross section is
\be
K_{0}\,(M^{2})\,
\frac{S}{s}\,\frac{\alpha_{s}T_F}{2\pi}
\left[
P_{qg}(z)\ln\left(\frac{M^{2}}{\mu^{2}}\right)
+ B(z)
\right]
+O(\eta),
\ee
where
\be
P_{qg}(z) = z^{2} + (1-z)^{2}
\label{eq:split1}
\ee
is the gluon $\rightarrow$ quark `splitting function', and
\be
B(z) = \left(\frac{1}{\eta}-\ln4\pi+\gamma_{E}\right)P_{qg}(z) +
P_{qg}(z)\,\ln\frac{(1-z)^{2}}{z} - \frac{3}{2}z^{2}+3\,z-\frac{1}{2}.
\ee
Denoting the appropriate product of quark and gluon
densities by $K(x_{1},x_{2})$ :-
\be
K(x_1,x_2) = (u(x_1)+\overline{d}(x_1))g(x_2)+
(u(x_2)+\overline{d}(x_2))g(x_1),
\ee
we can now write down the total $O(\alpha_{s})$ hadronic
cross section
\begin{eqnarray}
\sigma_{H} &=&
K_{0}\,(M^{2})\,
\int\frac{{\rm d}x_{1}}{x_{1}}\,\frac{{\rm d}x_{2}}
{x_{2}}H(x_{1},x_{2})\,
\left[\delta(1-z)+\theta(1-z)\,\frac{\alpha_{s}C_F}{2\pi}
\left(
2\,P_{qq}(z)\ln\left(\frac{M^{2}}{\mu^{2}}\right)
+ A(z)
\right)
\right]\nonumber\\
&&+\,K_{0}\,(M^{2})\,
\int\frac{{\rm d}x_{1}}{x_{1}}\,\frac{{\rm d}x_{2}}{x_{2}}
K(x_{1},x_{2})\,
\theta(1-z)\,\frac{\alpha_{s}T_F}{2\pi}\,\left[
P_{qg}(z)\ln\left(\frac{M^{2}}{\mu^{2}}\right)
+ B(z)
\right].
\end{eqnarray}
Expressing the `bare' quark densities in terms of their scale
dependent `renormalised' counterparts  we have
\begin{eqnarray}
q(x) &=& q(x,Q^{2}) -
\frac{\alpha_{s}C_F}{2\pi}\,\int_{x}^{1}\frac{{\rm d}z}{z}\,
\left(P_{qq}(z)\ln\left(\frac{Q^{2}}{\mu^{2}}\right)
+f_{q}(z)\right)\,q(x/z,Q^{2})\nonumber\\
&&-\frac{\alpha_{s}T_F}{2\pi}\,\int_{x}^{1}\frac{{\rm d}z}{z}\,\left(
P_{qg}(z)\ln\left(\frac{Q^{2}}{\mu^{2}}\right)
+ f_{g}(z)\right)\,g(x/z, Q^{2})+O(\alpha_s^2),
\label{eq:barequarkdensity}
\end{eqnarray}
where in the $\overline{\rm MS}$ factorisation scheme\cite{AEM}
\beqarr
f_{q}(z) &=& \left(\frac{1}{\eta}-\ln4\pi+\gamma_{E}\right)P_{qq}(z)
+(1+z^{2})\left(\frac{\ln(1-z)}{1-z}\right)_{+}
- \frac{3}{2}\,\frac{1}{(1-z)_{+}}\nonumber\\
&&-\left(\frac{9}{2}+\frac{1}{3}\pi^{2}\right)\,
\delta(1-z)-(1+z^{2})\,\frac{\ln z}{1-z}
+3+2z,\\
f_{g}(z) &=& \left(\frac{1}{\eta}-\ln4\pi+\gamma_{E}\right)P_{qg}(z)
+P_{qg}(z)\ln\frac{1-z}{z}+6\,z\,(1-z).
\eeqarr
Thus, with the removal of all bare parton densities we obtain
our result for the NLO hadronic cross section:
\begin{eqnarray}
\sigma_{H} &=&
K_{0}\,(M^{2})\,
\int\frac{{\rm d}x_{1}}{x_{1}}\,\frac{{\rm d}x_{2}}{x_{2}}
H(x_{1},x_{2};M^{2})\,
\left[\delta(1-z)+\theta(1-z)\,\frac{\alpha_{s}C_F}{2\pi}\,
2\,F^{(T)}(z)
\right]\nonumber\\
&&+\,K_{0}\,(M^{2})\,
\int\frac{{\rm d}x_{1}}{x_{1}}\,\frac{{\rm d}x_{2}}{x_{2}}
K(x_{1},x_{2};M^{2})\,
\theta(1-z)\,\frac{\alpha_{s}T_F}{2\pi}\,G^{(T)}(z)
+O(\alpha_s^2),
\label{eq:XSatoalphas}
\end{eqnarray}
where $F^{(T)}(z) = \frac{1}{2}A(z) -f_q(z)$
and $G^{(T)}(z) = B(z) - f_g(z)$ have the explicit forms
\begin{eqnarray}
F^{(T)}(z) &=& \frac{3}{2}\,\frac{1}{(1-z)_{+}}+
(1+z^{2})\,\left(\frac{\ln(1-z)}{1-z}\right)_{+} -2 -3\,z\nonumber\\
&&+\left(\frac{2}{3}\pi^{2}+\frac{7}{2}\right)\,\delta(1-z),
\label{eq:Ftot}\\
G^{(T)}(z) &=&
(z^{2}+(1-z)^{2})\ln(1-z)+\frac{9}{2}\,z^{2}-3\,z-\frac{1}{2}.
\label{eq:Gtot}
\end{eqnarray}
\par
One caveat in extracting numerical results from
eq.~(\ref{eq:XSatoalphas}) is that the renormalization scale of
$\alpha_s$ remains undetermined in our $O(\alpha_s)$ calculation.
Of course, this result could have been anticipated from the RGE
for $\alpha_s$:-
\be
\frac{{\rm d}\,\alpha_{s}}{{\rm d}\,t} \sim \alpha_{s}^{2},
\ee
where $\alpha_s^2 = 0$ to the level of approximation in
this calculation. In order to obtain numerical results one
must assume a scale for $\alpha_{s}$ and our choice,
although not rigorously justifiable without including $O(\alpha_{s}^{2})$
corrections, was the slepton mass `$M$'. The consequence of
choosing the the scale of $\alpha_s$ to be $S$ rather than $M$ is
simply to shift the cross section by terms of order $\alpha_s^2$.
We evaluated the NLO hadronic cross section (at $\sqrt{S} = 14$~TeV)
with several recent structure function sets using the PDFLIB
package\cite{PDFLIB}. Our principal result in Fig.~7 shows
the NLO hadronic cross section calculated using the
GRV~HO set\cite{GRVHO}  as a function of slepton
mass ranging from 100~GeV to 10~TeV.
\begin{center}
\makebox[352pt]{\epsfbox{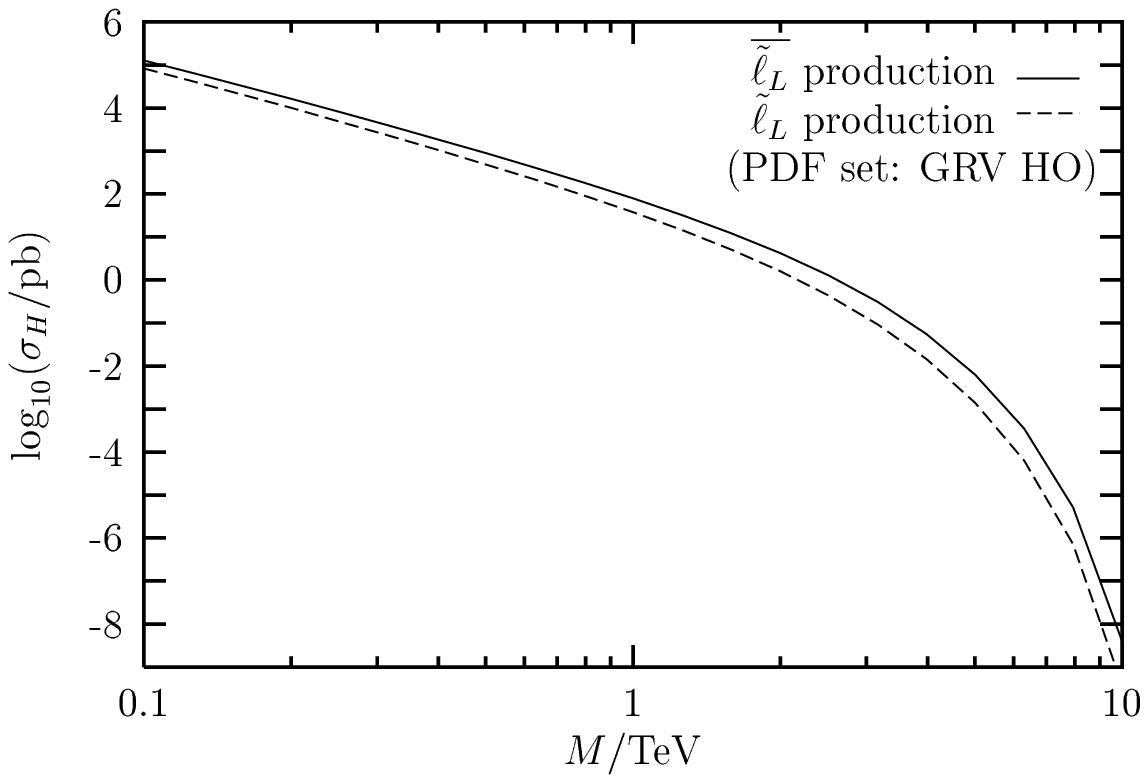}}\\
Figure~7:~The NLO hadronic cross section
\underline{for $\lambda_{i11}^{\prime} = 1$}\\
\end{center}
Fig.~8 shows the magnitude of NLO cross sections calculated
using three different PDF sets\footnote{All of the PDF sets used are NLL
evolutions calculated in the $\overline{\rm MS}$ factorisation
scheme and are consistent with current low-$x$ data.}
(namely, CTEQ 2pM\cite{CTEQ2pM}, MRS (H)\cite{MRSH} and
MRS D$^{\prime}_-$\cite{MRSD})
relative to the one in Fig.~7. In Fig.~9 we show the
ratio of the NLO to
tree level cross sections using all four PDFs. The fact that this
ratio increases with slepton mass simply means that the NLO cross
section falls off more slowly than the tree level cross section.
As one can see the corrections are quite significant, especially for
slepton masses in excess of $O(1)$ TeV.
\begin{center}
\makebox[334pt]{\epsfbox{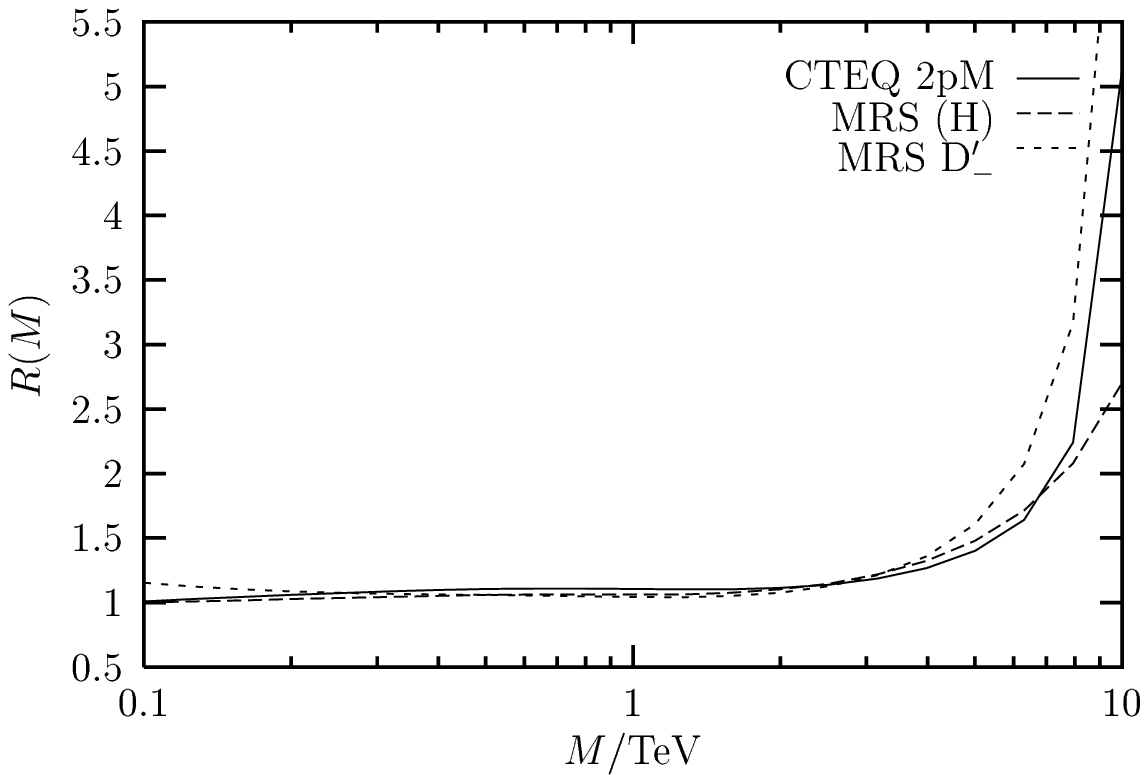}}\\
Figure~8:~$R(M) = \sigma_H^{(i)}/\sigma_H^{\rm (GRV~HO)}$;
$i=$ CTEQ 2pM, MRS (H), MRS D$^{\prime}_-$\\
\end{center}
\begin{center}
\makebox[347pt]{\epsfbox{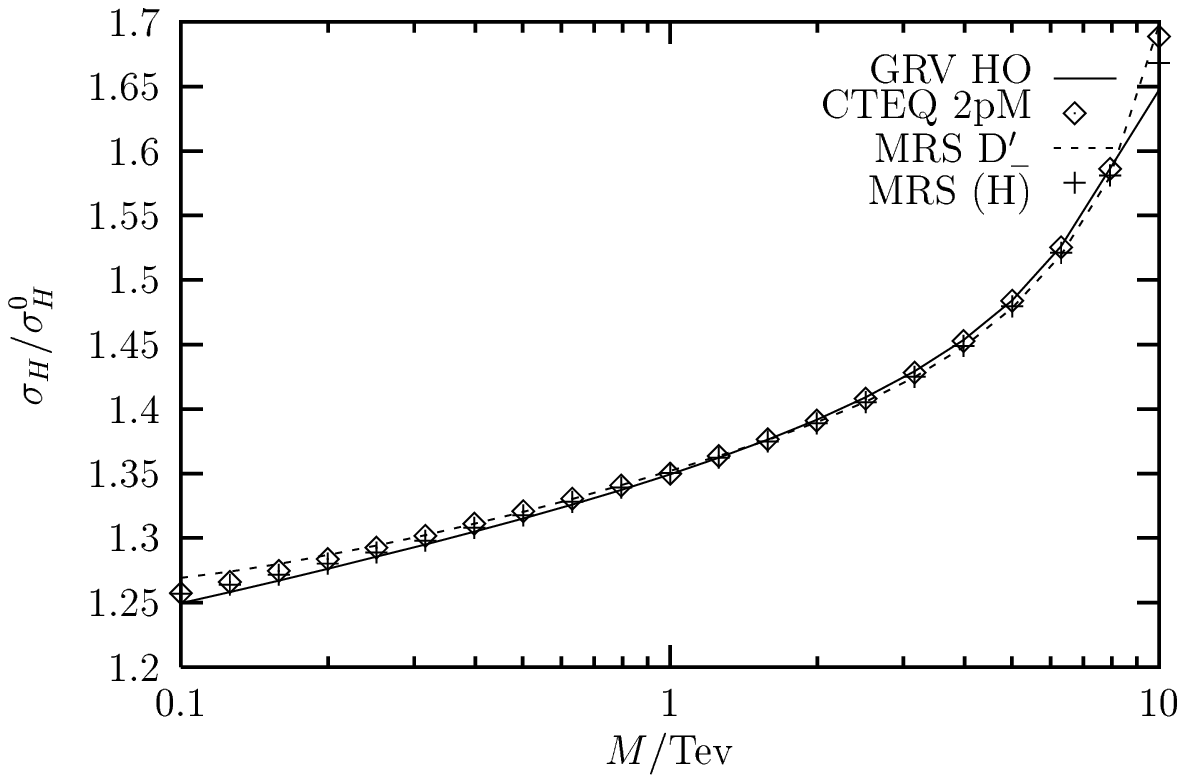}}\\
Figure~9:~Ratio of NLO to tree level cross sections\\
\end{center}
\par
Finally, we use the results displayed in Fig.~7 to determine the
critical coupling ($\lambda_c^{\prime}$) for `discovery' at the LHC.
Our results in Fig.~10 are based on a sample size of
$10^{5}$~pb$^{-1}$, (comparable to one year's worth of data)
and assume a discovery at 95\%CL. The
experimental bounds\cite{BGH} on $\lambda^{\prime}_{111}$, and
$\lambda^{\prime}_{211}$ ($\lambda^{\prime}_{311}$ is currently
unbounded) are plotted\footnote{These bounds are
valid if one assumes that the squarks and sleptons are degenerate}
for comparison on the same graph. As one can see, even in the worst
case, there is a window for discovery for $M\leq O(6)$~TeV.
\begin{center}
\makebox[330pt]{\epsfbox{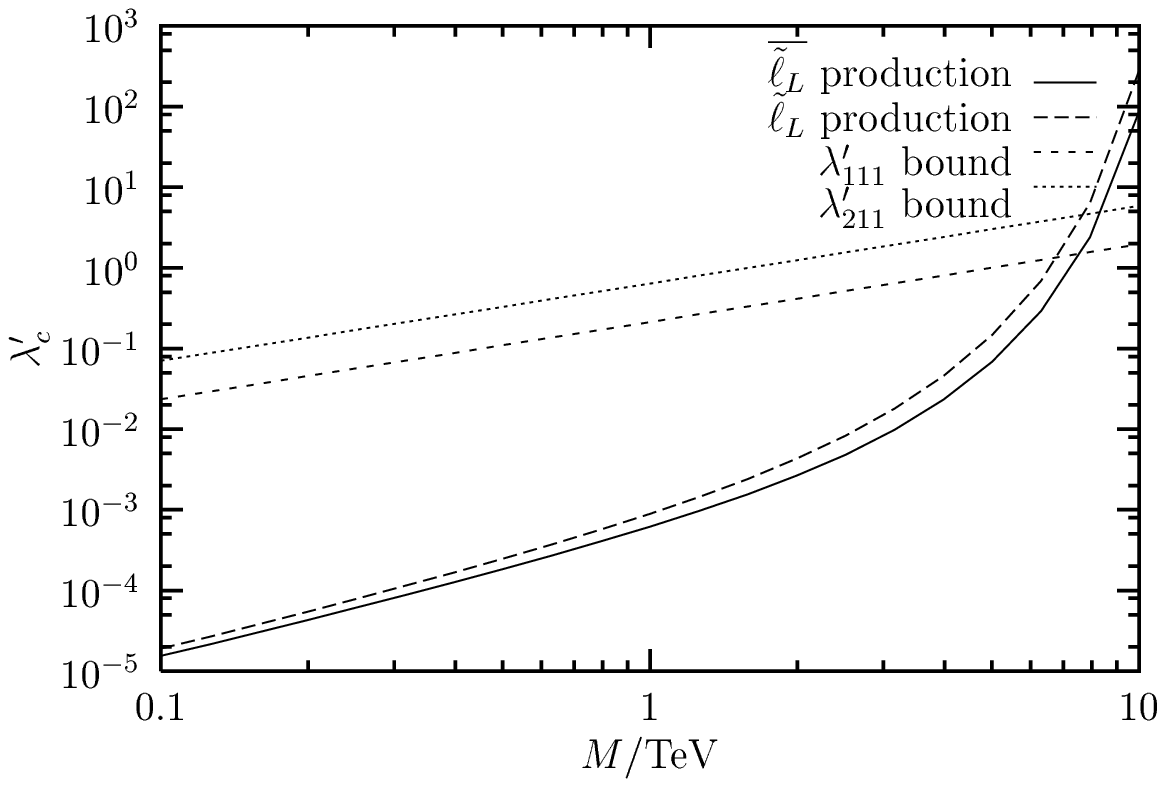}}\\
Figure~10:~Critical coupling for discovery in $10^5$~pb$^{-1}$ of data\\[1cm]
\end{center}
\section{Conclusion}
We have calculated the NLO hadronic cross section for monoslepton
at the LHC and have shown that the leading QCD corrections
are potentially large.
Depending on the ratio $\rho = m_{\tilde{\gamma}}/m_{\tilde{\ell}}$
the optimal signal for monoslepton production interpolates between a hard
isolated lepton with energy $O(m_{\tilde{\ell}}/2)$ recoiling
against a jet when $\rho \ll 1$, and an isolated like-sign dilepton
for $\rho \stackrel{<}{_\sim} 1$. Potential backgrounds from
$t \rightarrow b W^+$ or
$q\overline{q},\,H^0 \rightarrow W^+W^-$ can be excluded
by suitable energy cuts or by event topology.
\\[3ex]
{\bf Acknowledgements}\\
I would like to thank P. L. White and S. F. King for useful comments.
\\[1cm]
\noindent
\def\theequation{A\arabic{equation}}
\setcounter{equation}{0}
{\Large\bf
Appendix: QCD corrections in Landau gauge}
\\[2ex]
We now briefly present summarised results of a Landau gauge
calculation of the one loop QCD corrections. In this gauge we find
that the one loop QCD self energy correction
vanishes\cite{Marc}:
\be
\Sigma_{\rm QCD}(p) = 0.
\ee
Consequently, the only non-trivial contribution to $\Sigma(p)$ is from
$\Sigma_{\rm SUSY}(p)$ :-
\be
\Sigma(p) = -\frac{\alpha_s C_F}{4\pi}\,\not{\!p}\,
\left[\left(\frac{\overline{\mu}^2}{M^2}\right)^{\ve}
\frac{1}{\ve^{'}}
-2\int_0^1{\rm d}x \, x\ln\left( x(\xi^2_{\tilde{q}}
-\xi^2_{\tilde{g}})+\xi^2_{\tilde{g}}\right)
\right].
\ee
The resultant effect of the Landau gauge self-energy corrections is
to multiply the tree level cross section
by the overall factor
\be
\left[1-\frac{\alpha_s C_F}{2\pi}
\left(\frac{\overline{\mu}^2}{M^2}\right)^{\ve}
\frac{1}{\ve^{'}}\right]\,
\left[1+\frac{\alpha_s C_F}{2\pi}
\Delta_{\rm SE}(\xi^2_{\tilde{u}},\xi^2_{\tilde{d}},\xi^2_{\tilde{g}})
\right],
\ee
where
$\Delta_{\rm SE}(\xi^2_{\tilde{u}},\xi^2_{\tilde{d}},\xi^2_{\tilde{g}})$
is given in eq.~(\ref{eq:deltase}). Turning our attention to the
vertex correction we find
the contribution from the one loop QCD diagram in Landau gauge is the
multiplicative factor (c.f. eq.~(\ref{eq:finalqcdvert})):-
\be
\left[1+\frac{\alpha_s C_F}{2\pi}
\left(\frac{\overline{\mu}^2}{M^2}\right)^{\ve}
\frac{3}{\ve^{'}}\right]\,
\left[1+\frac{\alpha_s C_F}{2\pi}
\left(\frac{\overline{\mu}^2}{M^2}\right)^{\ve}
\left(-\frac{2}{\ve^2}-\frac{3}{\ve}+\frac{7\pi^2}{6}-2\right)
\right]+O(\alpha_s^2,\ve).
\ee
We now find the renormalization constants $Z_2$ and $Z_3$ to be
\beqarr
Z_2  = 1-\frac{\alpha_s C_F}{4\pi}
\frac{1}{\hat{\ve}},&&
Z_3  = 1-\frac{\alpha_s C_F}{4\pi}
\frac{3}{\hat{\ve}},
\eeqarr
and subsequently the Yukawa coupling renormalization constant $Z_1$ is
\be
Z_1 = \frac{Z_3}{Z_2} = 1-\frac{\alpha_s C_F}{2\pi}
\frac{1}{\hat{\ve}}
\ee
which, as expected, is identical to the result obtained in
the Feynman gauge.
\\[1cm]

\end{document}